\documentclass[aps,pra,twocolumn,showpacs,superscriptaddress,amssymb,floatfix,tightenlines]{revtex4}

\newcommand{\beq}{\begin{equation}}
\newcommand{\eneq}{\end{equation}}
\usepackage{graphicx}
\usepackage{amsmath}
\usepackage{epsfig}
\usepackage{dcolumn}
\usepackage{bm}
\input{epsf}

\newcommand{\ket}[1]{\left| #1 \right\rangle}

\newcommand{\braket}[2]{\langle #1 | #2\rangle}

\newcommand{\Tr}{\mathrm{Tr}}

\begin{document}

\title{Geometric multipartite entanglement measures}
\author {Gerardo~A. Paz-Silva}
\altaffiliation{E-mail address: {\tt gerapaz@univalle.edu.co}}
\affiliation{Departamento de F\'isica, Universidad del Valle, A.A. 25360, Cali, Colombia}
\author {John H. Reina}
\altaffiliation{Corresponding author. \\E-mail address:  {\tt j.reina-estupinan@physics.ox.ac.uk}; Tel/Fax: +57 2 3394610/3393237}
\affiliation{Departamento de F\'isica, Universidad del Valle, A.A. 25360, Cali, Colombia}
\affiliation{Institut f\"ur Theoretische Physik, Technische Universit\"at Berlin, Hardenbergstr. 36, 10623 Berlin, Germany}

\date{\today}

\begin{abstract}
Within the framework of constructions for quantifying entanglement, we
build a natural scenario for the assembly of multipartite entanglement
measures based on Hopf bundle-like mappings obtained through Clifford
algebra representations.
Then, given the non-factorizability of an arbitrary two-qubit
density matrix, we give an alternate quantity that allows the
construction of two types of entanglement measures based on their
arithmetical and geometrical averages over all pairs of qubits in a register of size
$N$, and thus fully characterize its degree and
type of entanglement. We find that such an arithmetical average is
both additive and strongly super additive.

\end{abstract}


%Quantum entanglement, 03.65.Ud
%Quantum information, 03.67.-a
%Quantum computation, 03.67.Lx

\maketitle

Quantum entanglement is arguably the most intriguing feature of the
quantum world, the hallmark of non-local correlations that draws the
line between classical and quantum behaviour~\cite{Bennett00}; it is
also the very resource that allows  quantum computation~\cite{comp},
and quantum protocols such as teleportation~\cite{Teleport} and key
distribution~\cite{crypto} to be performed. Since information can,
in principle, be processed by quantum technologies via the
manipulation of a given set of physical resources~\cite{Bennett00},
an important problem to be addressed is the quantification and
controlled manipulation of the degree of entanglement of a
continuously interacting physical system~\cite{Plenio}. In this
context, we should be capable of performing a reliable measure in
order to account for the quantification and processing of the
system's degree of quantum correlations at a given time. Many
efforts have been devoted towards this objective, ranging from
polynomial invariants~\cite{MeyerWallach,invariant,Emary}, density matrix
properties~\cite{density,Vedral,Plenio}, and positive
maps~\cite{Positive}, to geometrical/algebraic
approaches~\cite{Levay2004,jaegger,wooters}.

In this Letter we give a geometric formulation of multipartite
entanglement measures. In so doing, we construct a Hopf bundle like
mapping which allows us to obtain information about the entanglement
of a pure quantum state. The idea of using Hopf fibrations has been
explored by several authors in the context of the generalization of
Bloch spheres to higher dimensions~\cite{Levay,Mosseri,Bernevig}.
The two qubit case has an interesting interpretation in terms of the
second Hopf fibration~\cite{Levay}. The three qubit case, although
explored~\cite{Bernevig,Mosseri}, has not been fully understood.
Here we give the formulation for the  three qubit case and
generalize it to an $N$-qubit system. This implies a technical
challenge since  Hopf fibrations cannot be faithfully constructed in
higher dimensions. We propose a generalization in terms of Clifford
algebra representations, and build a Hopf-like map which gives
information about the system's entanglement. Moreover, we introduce
the permutation group to decode the information available to {\it
all} qubits in the system. This geometrical picture provides a
natural setting for the Meyer-Wallach entanglement
measure~\cite{MeyerWallach}. We also propose, to avoid some
inconveniences posed by the Meyer-Wallach measure, an alternate
strategy to quantifying multipartite entanglement. We introduce the
idea of probe quantities to characterize the non-factorizability of
each two-qudit density matrix within a pure quantum state and build
on it to construct two entanglement measures, that can also account
for the case of mixed states, by means of their
geometrical and arithmetical averages. We study their basic
properties and implications. We find that the proposed arithmetical
average is both additive and strongly super additive.

{\em Two-qubit entanglement and the second Hopf fibration}.---We
construct the two-qubit case following Ref.~\cite{Levay}, but fixing
the state norm to 1. The state vector can be written in the
computational basis as $\ket{\Psi} = \alpha_0 \ket{00} + \alpha_1
\ket{01} + \beta_0 \ket{10} + \beta_1 \ket{11}$, where $\alpha_0,
\alpha_1, \beta_0, \beta_1 \in \mathbb{C}$, and $|\alpha_0|^2 +
|\alpha_1|^2 + |\beta_0|^2 + |\beta_1|^2 =1$. Although there has
been a geometrical formulation of few-qubit entanglement in terms of
Hopf fibrations in Refs.~\cite{Mosseri,Bernevig,Levay}, this is
limited to only three Hopf fibrations which would impose a limit on
the system's scalability. Here we use a
 Hopf bundle-like mapping similar to the one reported in
 Ref.~\cite{Bernevig} but instead of using the geometrical picture
 of the spheres of the total space of the bundle we resort to the Clifford algebra
 representation. This
 construction allows us to avoid the limit imposed by Hopf
 fibrations and generalizes to higher dimensions. The relevance to a geometrical interpretation of
 states shall be dealt with later, as now we will emphasize its algebraic properties and build a map from the Clifford
$Cl(3)$ representation ($\tilde Cl(3) \sim
\mathbb{H}\oplus\mathbb{H}$) to $S^4$. We introduce the pair of
quaternions $q_1 = \alpha_0 + \alpha_1 i_2$, and $q_2 = \beta_0 +
\beta_1 i_2$. By choosing the map $\pi: \tilde Cl(3) \longrightarrow
S^4$:
\begin{eqnarray}
\nonumber
\ket{\Psi}   & = & (q_1,
q_2)\longmapsto \big(2q_2 \bar q_1, |q_1|^2 - |q_2|^2\big)   =  \\
\nonumber && \big( 2[\beta_0 \bar\alpha_0 + \beta_1\bar\alpha_1]+
2[\beta_1\alpha_0 -
 \beta_0\alpha_1] i_2, \,  \\
 &&  |\alpha_0|^2 + |\alpha_1|^2 - |\beta_0|^2 - |\beta_1|^2\big) \ ,
 \end{eqnarray}
 we access all the information
available from the first qubit and a component related to the so-called concurrence $\mathcal{C}$, a measure of the qubit system entanglement~\cite{Plenio}. Here $\bar x$ denotes complex conjugation.
Thus, the
map reads
\begin{equation}
\big(2\bar\rho_{1(01)} +\mathcal{C}_1 i_2,\, \rho_{1(00)} -
\rho_{1(11)}\big) \ ,
\end{equation}
where $\rho_{n(ij)}$ denotes the one-qubit reduced density matrix of
the $n$-th qubit. We point out that although we are keeping the map
for its algebraic features, it is manifestly invariant under local
SU(2) transformations and avoids the problem of defining charts as
in the Hopf bundle picture.  Such a map contains information about
the first qubit entanglement, which should be symmetrical
for the other qubit, and  the issue here is how to recover it. In the two
qubit case the other reduced density matrix can be recovered from
the first one simply by exchanging $\alpha_1$ with $\beta_0$. Of
course this is equivalent to labeling the qubits in the other possible
way, and in this sense it is a natural consequence of the arbitrariness
introduced by the label we imposed. For higher dimensions, however,
the number of possible labels is $N!$, so in principle we would have
$N!$ different mappings. We now introduce the permutation group as a
way of recovering the information for the other qubits. In fact, by
permuting the particles we obtain the desired exchange, or
equivalently, the other map:
\begin{eqnarray}
\nonumber
 \big(2(\bar\alpha_1 \alpha_0 + \bar\beta_1\beta_0) + 2(\beta_1\alpha_0 -
 \beta_0\alpha_1) i_2, \, |\alpha_0|^2 + |\beta_0|^2 -  \\ (|\alpha_1|^2 +
 |\beta_1|^2)\big) = \big(2\bar\rho_{2(01)} + \mathcal{C}_2 i_2, \, \rho_{2(00)} -
\rho_{2(11)}\big) \ . \, \, \, \, \, \, \, \,
\end{eqnarray}
As is to be expected, the term related to the concurrence is an
invariant: the entanglement of the two particles remains the same,
regardless of which one is measured. We will not go into the details
of discussing the interpretations of $\mathcal{C}$ (see e.g. Ref.~\cite{Levay}), instead
we shall focus on the fact that the map allows the construction of a
measure of entanglement  and that all possible maps are obtained
through the permutation group.

{\em Three-qubit entanglement and the third Hopf
fibration}.---Although the three-qubit scenario has been discussed
from the  point of view of Hopf fibrations ~\cite{Mosseri,Bernevig}, and
also from the perspective of the twistor geometry
formalism~\cite{Levay}, there is not a proper systematic
characterization regarding multipartite entanglement measures. Here
we provide such a description. We use the permutation group to
decode the information available from the Clifford algebra
representation. The Hilbert space for the three-qubit system is the
tensor product of the 1-qubit Hilbert spaces $\mathcal{H}_1 \otimes
\mathcal{H}_2 \otimes \mathcal{H}_3$ with the direct product basis
where a pure three-qubit state reads
$\ket{\Psi} = \alpha_0 \ket{000} + \alpha_1 \ket{001} +
\beta_0 \ket{010} + \beta_1 \ket{011} +
 \delta_0 \ket{100}
+ \delta_1 \ket{101} + \gamma_0 \ket{110} + \gamma_1
\ket{111}$,
$ \alpha_0, \alpha_1, \beta_0, \beta_1, \delta_0,
\delta_1,\gamma_0,\gamma_1$$\in \mathbb{C}$,  and $|\alpha_0|^2 +
|\alpha_1|^2 + |\beta_0|^2 + |\beta_1|^2 + |\delta_0|^2 +
|\delta_1|^2 + |\gamma_0|^2+ |\gamma_1|^2 =1$.
We perform a similar parametrization to the third Hopf fibration but
from the $Cl(4)$ representation to $S^8$:
\begin{eqnarray}
\nonumber
  \ket{\Psi}  & = &  (o_1, o_2) = (q_1 + q_2 i_4, q_3 + q_4 i_4); \, \, \, \,
q_1  =  \alpha_0 + \alpha_1 i_2, \\
q_2  & = &   \beta_0 + \bar\beta_1
i_2, \, \, q_3  =  \delta_0 + \delta_1 i_2,  \, \,
q_4  =  \gamma_0 +
\bar\gamma_1 i_2 \ ,
\end{eqnarray}
which results in the map $\pi': \tilde Cl(4) \longrightarrow S^8$:
\begin{eqnarray}
\label{mapeo1}
    \pi'(o_1,o_2) & = & ( 2  o_2 \bar o_1, |o_1|^2 - |o_2|^2) = \\ \nonumber
   &&  (2C_1+2C_2i_2+2C_3i_4+2C_4 i_6, |o_1|^2 - |o_2|^2)
\end{eqnarray}
with
$C_1=\bar\alpha_0 \delta_0 + \delta_1 \bar \alpha_1 +
\gamma_0 \bar\beta_0 + \bar\beta_1 \gamma_1$,
$C_2 = - \alpha_1 \delta_0 + \delta_1 \alpha_0 - (\bar \beta_1
\bar\gamma_0 - \bar\gamma_1
\bar\beta_0)$,
$C_3 = -\beta_0 \delta_0 + \gamma_0 \alpha_0 - ( \bar\alpha_1
\bar\gamma_1 - \bar\delta_1 \bar\beta_1)$, and
$C_4 = -\bar\delta_1 \bar\beta_0 + \bar\alpha_1 \bar\gamma_0 -
(\beta_1 \delta_0 - \gamma_1 \alpha_0)$,
where $|o_1|^2 - |o_2|^2 =\rho_{1(00)} - \rho_{1(11)}$, and
$1=|\braket{\Psi}{\Psi}|^4 =|\pi'(o_1,o_2)|^2$. Here $C_1$ is
clearly the off diagonal element of the reduced density matrix
$\rho_1$, but it is not trivial to see what exactly is the information
that is available from the other $C$'s. In the two-qubit case, we had
only one component, which was interpreted as the concurrence. In the three-qubit
scenario there is certainly more information to look at. Consider
the following qubit configurations:
i) $\ket{\Psi_1} \otimes \ket{\Psi_{23}}$,
$C_2=0$, $C_3=0$, $C_4=0$, ii)
$\ket{\Psi_{12}} \otimes \ket{\Psi_{3}}$,
$C_2=0$, $C_3 \neq 0$, $C_4\neq0$, iii)
$\ket{\Psi_2} \otimes \ket{\Psi_{13}}$,
$C_2 \neq 0$, $C_3=0$, $C_4\neq0$, and iv)
$\ket{\Psi_{1}} \otimes \ket{\Psi_{2}} \otimes \ket{\Psi_{3}}$,
$C_2=C_3=C_4=0$.
To some degree it may be tempting to say that for such configurations, the (123)
case as we shall call it from now on, $C_2$ has information about
the entanglement of (13) and $C_3$ of (12). An explicit  calculation reads:
i) case $1\otimes(23)$:
$2 C_1 = 2\bar\rho_{1(01)}$,
$2 C_2 = 0$,
$2 C_3 = 0$,
$2 C_4 = 0$,
ii) case $2\otimes(13)$:
$2 C_1 = 2\bar\rho_{1(01)}$,
$2 C_2 = 2a^2 C_{(13)} + 2\bar b^2 \bar {C}_{(13)}$,
$2 C_3 = 0$,
$2 C_4 = 4\, {\rm Im} (a b {C}_{(13)})$,
iii) case $3\otimes(12)$:
$2 C_1 = 2\bar\rho_{1(01)}$,
$2 C_2 = 0$,
$2 C_3 = 2a^2 C_{(12)} + 2\bar b^2 \bar {C}_{(12)}$,
$2 C_4 = 4 \, {\rm Re} (a b {C}_{(12)})$,
where $\mathcal{C}_{(13)}$ denotes the concurrence of the entangled
two-qubit state (13). It is clear that $C_2$ contains some
information about the entanglement of (13) and $C_3$ about (12). We
show next that, in contrast to the entanglement measure suggested in
Ref.~\cite{Bernevig}, this information is not complete and that it
is actually  the $N=3$ permutation group which indicates the
complete information available in the $Cl$ representation. We write
down the explicit transformations induced by the elements of the
permutation group of three elements as follows:\\
a) $(123) \rightarrow (213)$:
$\beta_1 \rightarrow \delta_1$, $\delta_0 \rightarrow \beta_0$, $\beta_0 \rightarrow \delta_0$, $\delta_1 \rightarrow \beta_1$, b) $(123) \rightarrow (321)$:
$\alpha_1 \rightarrow \delta_0$, $\beta_1 \rightarrow \gamma_0$, $\delta_0 \rightarrow \alpha_1$, $\gamma_0 \rightarrow \beta_1$,
c) $(123) \rightarrow (132)$:
$\alpha_1 \rightarrow \beta_0$, $\beta_0 \rightarrow \alpha_1$, $\delta_1 \rightarrow \gamma_0$, $\gamma_0 \rightarrow \delta_1$,
d) $(123) \rightarrow (312)$:
$\alpha_1 \rightarrow \delta_0$, $\beta_0 \rightarrow \alpha_1$,
$\beta_1 \rightarrow \delta_1$, $\delta_1 \rightarrow \gamma_0$,
 $\delta_0 \rightarrow \beta_0$, $\gamma_0 \rightarrow \beta_1$,
e) $(123) \rightarrow (231)$:
$\alpha_1 \rightarrow \beta_0$, $\beta_0 \rightarrow \delta_0$,
$\beta_1 \rightarrow \gamma_0$, $\delta_0 \rightarrow \alpha_1$,
 $\delta_1 \rightarrow \beta_1$, $\gamma_0 \rightarrow \delta_1$.
Hence, permuting the qubits is equivalent to
redefining five new mappings through the equivalences obtained above:
\begin{description}
\item  [$(123) \rightarrow (213)${\rm:}]
\item $C'_1 = \bar\alpha_0 \beta_0 +\beta_1\bar\alpha_1 + \gamma_0\bar\delta_0 + \bar\delta_1 \gamma_1$,
$C'_2 = \beta_1\alpha_0 -\alpha_1\beta_0 + \bar\gamma_1\bar\delta_0 - \bar\delta_1\bar\gamma_0$,
$C'_3 = \gamma_0\alpha_0 - \beta_0 \delta_0 -
(\bar\beta_1\bar\delta_1 - \bar\gamma_1\bar\alpha_1)$,
$C'_4 = - \beta_0 \delta_1 - \bar\gamma_0\bar\alpha_1 +
\bar\beta_1\bar\delta_0 + \gamma_1\alpha_0$,
$|o_1|^2 - |o_2|^2 = \rho_{2(00)} - \rho_{2(11)}$
\item [$(123) \rightarrow (321)${\rm:}]
\item $C'_1 = \bar\alpha_0\alpha_1 + \beta_1\bar\beta_0 + \delta_1\bar\delta_0 + \bar\gamma_0\gamma_1$,
$C'_2 = \delta_1\alpha_0 -\delta_0\alpha_1 +\bar\gamma_1\bar\beta_0 -\bar\gamma_0\bar\beta_1$,
$C'_3 = \beta_1\alpha_0 - \beta_0\alpha_1 + \bar\delta_1\bar\gamma_0 -\bar\delta_0\bar\gamma_1$,
$C'_4 =  - \gamma_0 \alpha_1 + \bar\delta_0\bar\beta_1 -
\bar\beta_1\bar\delta_0 + \gamma_1\alpha_0$,
$|o_1|^2 - |o_2|^2 = \rho_{3(00)} - \rho_{3(11)}$
\item [$(123) \rightarrow (132)${\rm:}]
\item $C'_1 = \bar\alpha_0 \delta_0 + \gamma_0 \bar \beta_0 +
\delta_1
\bar\alpha_1 + \bar \beta_1 \gamma_1$,
$C'_2 = \gamma_0\alpha_0 - \beta_0 \delta_0 -
(\bar\beta_1\bar\delta_1 - \bar\gamma_1\bar\alpha_1)$,
$C'_3 = \delta_1 \alpha_0 - \delta_0 \alpha_1 -
(\bar\beta_0\bar\gamma_1 - \bar\gamma_0\bar\beta_1)$,
$C'_4 = \bar \beta_0 \bar\delta_1 - \bar\gamma_0\bar\alpha_1 -
(\beta_1\delta_0 - \gamma_1\alpha_0)$,
$|o_1|^2 - |o_2|^2 = \rho_{1(00)} - \rho_{1(11)}$
\item [$(123) \rightarrow (312)${\rm:}]
\item $C'_1 = \bar\alpha_0 \beta_0 +\beta_1\bar\alpha_1 + \gamma_0\bar\delta_0 + \bar\delta_1 \gamma_1$,
$C'_2 = \gamma_0\alpha_0 - \delta_0\beta_0 + \bar\gamma_1\bar\alpha_1 - \bar\delta_1\bar\beta_1$,
$C'_3 = \beta_1\alpha_0 - \alpha_1\beta_0 + \bar\gamma_0 \bar\delta_1 - \bar\delta_0\bar\gamma_1$,
$C'_4 = - \beta_0 \delta_1 - \bar\gamma_0\bar\alpha_1 +
\bar\beta_1\bar\delta_0 + \gamma_1\alpha_0$,
$|o_1|^2 - |o_2|^2 =\rho_{2(00)} - \rho_{2(11)}$
\item [$(123) \rightarrow (231)${\rm:}]
\item $C'_1 = \bar\alpha_0\alpha_1 + \beta_1\bar\beta_0 + \delta_1\bar\delta_0 + \bar\gamma_0\gamma_1$,
$C'_2 = \beta_1\alpha_0 - \beta_0\alpha_1 +\bar\gamma_1\bar\delta_0
-\bar\gamma_0\bar\delta_1$,
$C'_3 = \delta_1\alpha_0 -\delta_0\alpha_1 +\bar\beta_1\bar\gamma_0 - \bar\beta_0\bar\gamma_1$,
$C'_4 = \bar\beta_0 \bar\delta_1 - \gamma_0\alpha_1 -
\bar\beta_1\bar\delta_0 + \gamma_1\alpha_0$,
$|o_1|^2 - |o_2|^2 =\rho_{3(00)} - \rho_{3(11)} \ .$
\end{description}
From this we see that (123) and (132) have information about the
first qubit, (213) and (312) about the second qubit, and (321) and
(231) about the third qubit. Although the density matrix information
is the same for (123) and (132), the information of the entanglement
available in (123) is not the same as that of
(132)~\footnote{However, an invariant common to both can be
constructed using the norm of the vector.}, and this is why the use
of the \textit{complete permutation group} of $N = 3$ particles is
what gives the complete information available from the tripartite
system. For higher dimensions, and once a map such as
Eq.~\eqref{mapeo1} is constructed, permutations will always yield
similar maps with the first and last components related to the
reduced density matrices of the system. A natural way of extracting
the information hidden in the parametrization is given through the norm of
the elements of the base space. We thus define the quantity $K_1$ as
\begin{align}
\begin{split}
\label{primvez} K_1 &= |2 C_2|^2 + |2 C_3|^2 +|2 C_4|^2\\
&= 1 - |2C_1|^2 - ||o_1|^2 -
|o_2|^2|^2\\
&= 1 - 4|\rho_{1(01)}|^2 - |\rho_{1(00)} - \rho_{1(11)}|^2 ± \ ,
\end{split}
\end{align}
which, after some manipulation, is equivalent to
\begin{equation}
\label{MW} K_1 = 2 ( 1 - \Tr[\rho^2_1])\ .
\end{equation}
It is encouraging that this coincides with the Meyer-Wallach-Brennen
quantities defined to measure entanglement in Ref.~\cite{Brennen}, which can be seen by
rewriting the measure built by Meyer and
Wallach~\cite{MeyerWallach} through an analysis of invariant
polynomials and the linear entropy. Indeed, an
arithmetical average over all permutations yields the
Meyer-Wallach-Brennen measure
\begin{align}
\begin{split}
\label{meycon}  \tilde M &= \frac{1}{N!} \sum^{N!}_1 K_i
= \frac{1}{N!} \sum^{N!}_1 2( 1 - \Tr[\rho^2_i]) = \\
&= \frac{1}{N} \sum^{N}_1 2(1 - \Tr[\rho^2_i])
\equiv \frac{1}{N} \sum^{N}_1 Q_i \ ,
\end{split}
\end{align}
where $\rho_i$ are the reduced density matrices of one qubit
accessible through the action of the permutation group. In the last equality we have used the fact that each $K_i$
($\rho_i$) appears with multiplicity $N-1$. For higher dimensions we
can build similar maps to Eq.~\eqref{mapeo1} using higher Clifford
algebra representations in such a way that
\begin{eqnarray}
\nonumber
&& |2C_2|^2+|2C_3|^2+|2C_4|^2+\hdots + |2C_{2^{n-1}}|^2   =   \\  && 1 - |2\bar\rho_{1(01)}|^2 -
|\rho_{1(00)}-\rho_{1(11)}|^2  \ .
\end{eqnarray}
{\em Multipartite qubit entanglement}.---The case of dimension $N \geq 3$ requires special
attention. It is known that the Meyer-Wallach-Brennen measure is not at all successful at   accurately measuring the entanglement when
$N\geq 3$.  The first qualitatively different scenario we find is the
$N=4$ case~\footnote{For higher $N$ we would have different
partitions, e.g. $6 = 2+2+2 = 3+3=...$ etc.}, which allows the
additional possibility of being factorized as a (2+2)-qubit state.
In particular, the case of the direct product of two EPR states
gives $\tilde M =1$, which appears as an unexpected result since
this is a semi factorizable state. This is so because, for such
states, the corresponding reduced density matrices are equal to
$\frac{1}{2} \hat I$ and hence the information about the
entanglement available for each qubit is the same: maximal
entanglement.

Scott~\cite{Scott} has addressed this issue by  considering not only
one qubit reduced density matrices, but the reduced density matrices
of $m<N$ qubits, corresponding to other bipartitions: 2 + 2, 3 + 2,
etc. He has proposed  the measure
\begin{align}
\begin{split}
  Q_m(\Psi) &= \begin{pmatrix} N \\ m \end{pmatrix}^{\!\!-1}
  \sum_{|S|=m}\frac{2^m}{2^m-1}\left(1-\Tr\Big[\rho_S^2\Big]\right)  \ ,
\end{split}
\label{Q}
\end{align}
where $ C^N_m = \begin{pmatrix} N \\ m \end{pmatrix}$. The choice $N
= \lfloor N/2 \rfloor$ would account for all possible bipartitions,
and as any $s$-partition is necessarily included in a bipartition,
then it should be enough just to consider bipartitions. In a similar
spirit, Love {\em et al.}~\cite{Multientanglement} proposed a
geometrical average of the term over which the sum is performed in Eq.~(\ref{Q})  as a way of
characterizing global entanglement. Nevertheless, these type of
measures have the inconvenience that they do not yield one for
generalized GHZ states, as all their reduced density matrices have
purity equal to 1/2 yielding a value equal to one only for the case
of two qubit reduced density matrices.

Although this can, in principle, be overcome by introducing a proper
normalization factor, we seek an alternate measure that can be
averaged in a more satisfactory way, namely, that yields one for GHZ
states and gives a non-zero value for maximally mixed density
matrices. Our proposal is that the arithmetical and geometrical
averages of such a quantity are enough to characterize the
entanglement of a quantum state, quantifying it and deciding whether
the entanglement is global (among all parties) or not
(bifactorizable, ..., $N$-factorizable).

Based on the factorizability of a given density matrix, we propose
an alternate strategy which is both economic and effective at
quantifying multipartite entanglement. The basic idea is to
characterize the degree of entanglement of a system through a probe
quantity, say  $\mathcal{P}$, that measures the degree of
non-factorizability ($\rho_{AB} = \rho_A \otimes \rho_B$) of each
pair of qudits.

Note that factorizability is only equivalent to separability in the case of pure density matrices. For the case of mixed density
matrices, a factorizable density matrix $\rho = \rho_A \otimes \rho_B$ can indeed be rewritten as a separable matrix $\sum p_i \rho^i_A \otimes \rho^i_B$, however, the converse is not true (see e.g. the Werner states). We focus on qubit systems ($d=2$).

First, we have to find a suitable
candidate for this task. Several candidates can be thought of, e.g.,
the mutual information~\cite{Cerf-MI}, the trace distance, and
others~\cite{comp}.  For the sake of completeness in our analysis
we consider two main quantities: i) The first quantity, which we term as the {\it
quasi-concurrence} $\mathcal{Q}_C(\rho_{AB})$, is based in the same eigenvalues $\lambda_i$ of the
concurrence \cite{wooters}, but considers a different combination to that defined in Ref. \cite{wooters}. Thus, we consider,  in
decreasing order, the eigenvalues $\lambda_i$ of the matrix
$\sqrt{\rho_{AB} \tilde\rho_{AB}}$, where $\tilde\rho_{AB} =
(\sigma_2\otimes\sigma_2) \bar\rho_{AB} (\sigma_2\otimes\sigma_2)$, in order to define
$\mathcal{Q}_C(\rho_{AB})=  \lambda_1 + \lambda_2 - \lambda_3-\lambda_4$. Note
that $\mathcal{Q}_C$ is non-negative, ranging from zero to one, and that for pure
states is equivalent to the concurrence $C(\rho_{AB})= \max
\{0,\lambda_1 -\lambda_2-\lambda_3-\lambda_4\}$, as we only have one
non-zero eigenvalue. Also, $\mathcal{Q}_C(\rho_{AB})=0$ for factorizable density matrices,
since all their eigenvalues are the same; and all the
reduced density matrices within a generalized GHZ state $\ket{GHZ} = (1 /
\sqrt{2})(\ket{0}^{\otimes N}+\ket{1}^{\otimes N})$, yield
$\mathcal{Q}_C(\rho_{AB}) = 1$, thus being a well suited quantity for our purposes.
ii) The second quantity, $\mathcal{F}r(A,B)$, has a more direct physical significance. This is
proportional to the von Neumann's mutual information
\begin{eqnarray}
\mathcal{F}r(A,B)  &=& \frac{1}{2}
\Big(S(\rho_{A})+S(\rho_{B})-S(\rho_{AB})\Big),
\end{eqnarray}
where $S(\rho)$ is the von Neumann's entropy associated to the
density matrix $\rho$~\cite{comp}.  For example,  for a generalized
 state $\ket{GHZ}$, all the two qubit reduced density matrices
yield $\mathcal{F}r(A,B) = 1/2$; a W-state $\ket{W} = (1/ \sqrt{3})
(\ket{100}+\ket{010}+\ket{001})$ has $\mathcal{F}r(A,B)\sim 0.46$
for all $(A,B)$, and a fully factorizable state yields
$\mathcal{F}r(A,B) =0$.

Second, we define geometric and arithmetic
averages in order to obtain  more information about the state and
its type of entanglement. We define the arithmetic average
\beq \label{arit} \mathcal{M} = \mathcal{N}(\mathcal{P}(A,B))
\big(C^N_2\big)^{-1}\sum \mathcal{P}(A,B) \ , \eneq
 and the geometric average
 \beq
\label{geom} \mathcal{G} =  \mathcal{N}(\mathcal{P}(A,B))\Big(\prod
\mathcal{P}(A,B)\Big)^{\big(C^N_2\big)^{-1}} \ , \eneq
where $\mathcal{P}(A,B)$ are the probe quantities: $\mathcal{Q}_C(\rho_{AB})$ and
$\mathcal{F}r(\rho_{AB})$ in our case, and the sum and product are over all
possible pairs of qubits. We note that
$\mathcal{P}{(A,B)}=\mathcal{P}{(B,A)}$ and that the
normalization factor, $ \mathcal{N}(\mathcal{P}(A,B))$, is introduced so the measure
yields one for generalized GHZ states. We have $\mathcal{N}=(1 +
(d-1)(1-\delta_{2,N}))$ if we are averaging $\mathcal{F}r$ or
$\mathcal{N}=1$ if we are using the quasi-concurrence.

The generalization to the case of mixed states can be achieved
through the expressions~\cite{Plenio}
\beq \mathcal{M}(\rho) = \min \sum p_i \mathcal{M}(\rho_i) \ ,
\eneq
and
\beq \mathcal{G}(\rho) = \min \sum p_i \mathcal{G}(\rho_i) \ ,
\eneq
for the arithmetical and geometrical averages respectively. The
minimum is intended over all possible decompositions. Note that in
the two-qubit case, $\mathcal{M}(\rho) = \mathcal{G}(\rho)$ and they
reduce to the concurrence defined by Wooters \cite{wooters}. It
is important to note that the probe quantities are directly applied without minimizations when
they are acting as probes on a multipartite quantum state; when
analyzing bipartite states, however, they act as full measures so they
operate through minimizing mechanisms such as the ones given above.

Both quantities give different information about the type of
entanglement of the quantum state, namely if it has global
entanglement (among all parties) or only among some of the parties
(bipartitions, tripartitions, etc.), and, in conjunction, they give us the
possibility of fully characterizing the entanglement of a pure quantum
state.

The arithmetic average varies between zero and one: it equals zero
if and only if all $\mathcal{Q}_C(\rho_{AB})$'s (or
$\mathcal{F}r{(A,B)}$'s)  are zero, that is, if the state is
completely factorizable, and equals one if and only if all
$\mathcal{Q}_C(\rho_{AB})$'s are 1 (or $\mathcal{F}r{(A,B)}$'s are
1/2), i.e. if we have an $N$-qubit maximally entangled state. It can
be shown~\cite{PazReina2-2006} that the measure defined in this way
is both additive and strongly super additive, two properties that
are desired but usually not satisfied by  most entanglement
measures~\cite{Christandl-TC}. Thus, we provide a measure of
entanglement that accurately quantifies the amount of entanglement
of a pure quantum state. This measure is on its own, however, unable
to determine whether a state is genuine globally entangled or not.
For example, a bi-factorizable $N$-qubit state could yield the same
value as a globally entangled state.

The geometric average also varies between zero and one. This equals
one if and only if all $\mathcal{Q}_C(\rho_{AB})$'s
($\mathcal{F}r{(A,B)}$'s) are 1 (1/2), that is, if we have a
maximally entangled $N$-qubit state. However, and  in contrast with
the arithmetic measure, it equals zero if at least one of the
$\mathcal{P}{(A,B)}$ vanishes, i.e., if the state is at least
bifactorizable, thus quantifying {\it global} entanglement. Higher
factorizabilities, e.g. trifactorizable or $N$-factorizable states
would have even more vanishing $\mathcal{P}{(A,B)}$'s.

These two measures lead us to classify the set of pure multipartite
states between the ones with non vanishing $\mathcal{G}$, i.e.
genuine globally entangled states\cite{PazReina2-2006}, and the
rest, which may or not be entangled at all, information that can be
obtained through  the arithmetical average $\mathcal{M}$. We recall
that according to our definition a state is genuine globally
entangled if after measuring one qubit we gain some information
about all of the other qubits in the register, that is if
$\mathcal{P}(A,B)\neq 0$ for all $(A,B)$. Both measures provide a
faithful way of discriminating between pure multipartite quantum
states. We can further explore the structure of the globally
entangled states; there are two distinct types: the ones with the
same $\mathcal{P}{(A,B)}$ for all pairs of qubits (homogeneously
entangled states) and the states that posses different values of
non-factorizability for all their pairs (heterogeneously entangled
states). In the context of this classification, our calculations
lead us to define a GHZ state as the homogeneously entangled state
with the highest possible average of $\mathcal{P}{(A,B)}$.

We introduce this definition because, if we consider the case of the
quantity $\mathcal{F}r$ \footnote{If, instead of $\mathcal{F}r$, we
consider  the quasi-concurrence, the definition is even more
compact; it would then read:  a GHZ state is that for which
$\mathcal{P}{(A,B)}$ is maximum for all reduced bipartite density
matrices.}, given an arbitrary state we may find higher values of
correlations between a pair of qubits, e.g., i) $\ket{EPR} \otimes
\ket{EPR}$ states, and ii) maximally entangled mixed states
\cite{MEMS}, but this does not mean that their degree of
entanglement is higher than that of the $\ket{GHZ}$ state. For the
case i) $\mathcal{F}r{(A,B)} = 1 > 1/2$ for $(A,B) = (1,2),(3,4)$,
and $\mathcal{F}r{(A,B)} = 0$ for $(A,B) = (1,3),(1,4),(2,3),(2,4)$,
yielding $\mathcal{M} = 2/3 < 1 = \mathcal{M}_{GHZ}$. Note that
$\mathcal{G} = 0 < 1 = \mathcal{G}_{GHZ}$. For the case ii) consider
the pure state $\ket{MEMS} = \sqrt{1-x} \ket{0101} + \sqrt{x}/2
(\ket{0000} +\ket{0011}+\ket{1100}+\ket{1111})$, built as a
purification of the maximally mixed entangled state proposed by
Munro {\it et al.} \cite{MEMS}: a direct calculation yields
$\mathcal{F}r{(1,2)} = \mathcal{F}r{(3,4)}  > 1/2 =
\mathcal{F}r{(1,2)}(\rho_{GHZ})$ for $x \rightarrow 1$. Despite this
result, it is not difficult to check that $\mathcal{F}r{(1,3)} =
\mathcal{F}r{(1,4)} = \mathcal{F}r{(2,3)} =\mathcal{F}r{(2,4)} <
1/4$ which gives $\mathcal{M} < \mathcal{M}_{GHZ}$, and also
$\mathcal{G}  < \mathcal{G}_{GHZ}$. In this way we see that,
although these states exhibit a higher degree of non-factorizability
among some of their pairs, on average they possess less entanglement
than the GHZ state, which possesses the maximum possible
entanglement on average. It is interesting that this result implies
the possibility that weaker (less correlated) links may be taking
place in spin chains or related systems~\cite{Bose}. 

We presented a geometrical formalism which is suited to describing
the Meyer-Wallach~\cite{MeyerWallach} measure in its
Brennen's~\cite{Brennen} version, thus generalizing previous work by
 Bernevig~\cite{Bernevig}, Mosseri~\cite{Mosseri}, and
Levay~\cite{Levay}. We also introduced, through the concept of probe
quantities characterizing the non-factorizability of a bipartite
density matrix, the construction of arithmetic and geometric
entanglement measures for quantifying multipartite pure states,
which were able to distinguish between globally and partially
entangled states and that accurately quantified the degree of
entanglement of an $N$-qubit system.

{\bf Acknowledgments}.
We  thank  T. Brandes, C. Emary, and H. Ocampo for useful discussions.
JHR thanks S. Reina-Steers for useful distractions. GAPS thanks G. H. Paz,
I. Silva, D. F. Gutierrez, and G. R. Paz for continual support.
We gratefully acknowledge financial support from COLCIENCIAS under
Research Grants No. 1106-14-17903 and No. 1106-05-13828.

\end{document}